# A Puzzling "Mule" Coin from the Parabita Hoard: a Material Characterisation

G. Giovannelli (1), S. Natali (1), B. Bozzini (2), D. Manno (3), G.Micocci (3), A. Serra (3),G. Sarcinelli (4), A. Siciliano (4), R. Vitale (4)

((1) Dipartimento ICMMPM, Università di Roma "La Sapienza", Roma, Italy. (2) Dipartimento di Ingegneria dell'Innovazione, Università di Lecce, Lecce, Italy. (3) Dipartimento di Scienza dei Materiali, Università di Lecce, Lecce, Italy.(4) Dipartimento dei Beni Culturali, Università di Lecce, Lecce, Italy.)

ABSTRACT

In this research, we report on the compositional, microstructural and crystallographic properties of a lead coin which has been regarded for many years as a genuine silver coin minted in the Southern Italy in the course of the 4th century BC. The material characterisation of this object allowed detecting an ancient forging technology, not previously reported, which was meant for the silvering of lead substrates The data collected have disclosed a contemporary counterfeiting procedure based on a metal coating process onto a Pb substrate. This coating has been identified as a bi-layer with a Cu innermost and an Ag outermost visible layer. As far as the coating application technique is concerned, the gathered evidence has clearly indicated that the original appearance of this artifact cannot be explained in terms of any of the established methods for the growth of an artificially silvered coating in classical antiquity. This technology is now being explained in terms of modern, fully non destructive scientific methods.

1. INTRODUCTION

In this study, a lead coin that is part of a monetary hoard unearthed in 1948 near Parabita (LE), now on display at the Taranto MNA, has been subjected to material characterisation.. This artifact (entry number 13 in the inventory list for the Parabita hoard) is catalogued as item #760 in the Fischer-Bossert monograph on the silver nomoi of Tarentum [Fischer-Bossert 99]. Rather surprisingly, the elemental composition of this coin went undetected through several examinations, until the last few years. It was not until 2003 that the EDX data collected in the course of a comprehensive research, performed at DBC of Lecce, on the silver content of coins from the Greek colonies in Southern Italy, allowed the identification of lead as the minting metal for this specimen.

This achievement raised far more questions than answers, inasmuch as the recognition of a monetary role of metal lead in classical antiquity remains a controversial issue within the numismatic literature [Siciliano 93]; much of the controversy is concerned with whether to assign a deceptive character to these objects, just as in the case of the surface silvered copper coins mimicking the coin types of regular issues, or not. This uncertainty (which may depend on the paucity of written, archeological, hoard, and die-study evidence on the coin–like lead pieces which are seldom recovered from archaeological contexts) is reflected in the terminological scatter among the scholars who tackled the problem: pseudo-coins, cast coins, lead tessaræ and lead die trials [Siciliano 93].

As far as the "deceptivity issue" is concerned, subplumblean coins are attested by the Athenian statesman Demosthenes and an Athenian law inscription of the $4^{th}$ century BC, so there is no doubt that counterfeit lead coin did exist, although the normal way to counterfeit coins was to use a bronze core [La Niece 93]. As it will be shown in the following sections, the material characterisation of this object allowed detecting an ancient forging technology, not previously reported, which was meant for the silvering of lead substrates; this technology is now being explained in terms of modern, fully non destructive scientific methods.

The coin is identified by [Fischer-Bossert 99]. as an *unicum* because of its being minted with mismatched sides that combine obverse and reverse designs, both pertaining to the Western Greek coinage [Giovannelli 05], but not usually paired in regular issues. The reverse of the coin is linked (typologically, but not technically) to a series of Tarentine didrachms minted in the years 333-331/0 b.C., the standard weight of which is 7.8 g. The obverse design of the coin belongs to a series of so-called Campano-Tarentine didrachms [Stazio 80], minted in the middle of the $3^{rd}$ century b.C., not in Taranto, but presumably in Campania or Lucania. The standard weight of this latter series is about 0.8 g lighter than the Tarentine one of the reverse. The specimen Parabita 13 is 8.05 g and consequently, as weight considerations dictate, it was attributed by [Fischer-Bossert 99] to the heavier (i.e. older) series.

2. EXPERIMENTAL

*2.1 Morphological characterisation*
The morphology of the sample was studied with a Cambridge Stereoscan 360 SEM SEM. The electron source was $LaB_6$. Electron detection was carried out with a scintillation photodetector. The typical working pressure was $10^{-7}$ mbar. Macrophotography images were obtained on a Nikon Coolpix 3500.

*2.2 Structural characterisation*
Structural characterisation was performed by XRD, using a Philips PW 1830 diffractometer equipped with a Philips PW 1820 vertical Bragg-Brentano powder goniometer and a Philips 1710 control unit. The scan rate adopted was 1 deg/s. The employed radiation was unmonochromated Cu K$\alpha$. Generation of structure models and theoretical diffractograms calculations were accomplished using the PowderCell (site http://www.bam.de/service/publikationen/powdercell_i.htm) software.

*2.3 Chemical characterisation*
Quantitative analysis of the sample composition was performed by EDX in the same vacuum chamber and with the same electron source as SEM with a Li-doped Si detector. Compositional data have been collected also by means a dedicated ED-XRF spectrometer (Spectro X-test, lateral resolution of 1x1 $mm^2$, penetration range about 20 μm), equipped with a Cu anode and a light element detector, only suitable for qualitative analysis

## 3. ANALYSIS OF THE COIN
### 3.1 Morphological and compositional analyses

This coin did not receive any conservation treatment and consequently it has been subjected to the investigation described below in as-found conditions. As illustrated in thumbnail images shown in Figures 1 and 2 it is almost completely covered with a greenish layer of corrosion products; on both sides of the coin the peripheral area is sparsely covered with the remnants of what could have been an applied white metal coating. Digital image processing and zooming on selected areas of Figure 2, allowed the detection of some subtle features that would go unnoticed in un-enhanced imagery:

(a) On the uncovered area of the coin (Figure 3) some surface flaws recalling the solidification patterns of a cast metal core [Devoto 93] are observed; these flaws are partially obliterated in the relief portions by intentional brushing (two nearly orthogonal bands of roughly parallel carved micro-glyph can be detected on the device of the coin. Intentional brushing (Figure 4) is also observed on the rim of the artefact seemingly to eliminate possible edge seams.

(b) Figure 5, showing some surface scratches (may be purposely inflicted in order to provide evidence of forgery) on the remnant coating strongly suggest that this could be constituted by a two-metals layer.

The SEM micrograph of the residual white-metal coating, reported in Figure 6, exhibits a cauliflower granular morphology with average diameter of ca. 15÷20 μm; such morphologies are typical of metals grown by electrochemical processes. The tips of the granuli are worn and exhibit a strong compositional contrast with respect to the rest of the crystallite. EDX mapping shows that the material of the worn tip of the grains is Cu while rest of the coated surface is Ag. These facts support the interpretation of the coating as a bi-layer with a Cu underlayer and an Ag top-layer.

The EDX spectra of the uncovered areas showed clear Pb, Sb, O and C peaks, thus hinting at an heavily corroded inner core of a Pb rich (90-95% Pb) Pb-Sb alloy. XRF spectroscopy confirms that the metal content of the uncovered areas is Pb and Sb.

### 3.2 Structural analyses

The topics of the alteration products on lead metallic archaeological findings has been covered in [Guida 80] and the measurement of the corrosion extent of archaeological lead artefacts has been performed by [Reich 03]; according to these studies, plattnerite ($PbO_2$), cerussite ($PbCO_3$), the orthorombic (massicot) and the tetragonal variety (litharge) of PbO, cotunnite ($PbCl_2$), phosgenite ($Pb_2CO_3Cl_2$), anglesite ($PbSO_4$) and hydrocerussite ($Pb_3(CO_3)_2(OH)_2$) are expected. Figure 7 shows an X-rays diffractogram of the uncoated portion of the coin. Indexing of this XRD pattern reveals that the surface area is composed of Pb corrosion products (mainly hydrocerussite, massicot and litharge) with metallic Cu and Pb as minor components. No Ag-related reflections could be detected. The XRD pattern shown in Figure 9, pertaining to the coated area, indicates the presence of metallic Ag, Cu and Pb (the latter possibly deriving from the substrate), along with the same mineral alteration products of Pb found in the uncoated portion of the coin. The texture analysis of the reflections pertaining to both Ag and Cu indicate a (111) preferred orientation, which can hardly be reconciled with a pyrometallurgical coating process, as it would be the case for a subærati-type counterfeit coin [Bozzini 03]. In fact, the diffractograms of hammered-and-annealed metals [Taylor 61] would show (220) and (311) as the most intense reflections, this ensuing from the very complex mixed textures that fcc metals develop upon thermomechanical treatments [Dong 95], [Pinol 01]. The observed (111) prevailing orientation is quite surprisingly typical of an electrochemical deposition process [Popov 02].

4. CONCLUSIONS

As long as nobody proves that this coin is a modern intrusion into the Parabita hoard, the collected data reveal a contemporary counterfeiting procedure (most probably dating back to the 3$^{rd}$ century b.C. when combing types of two different series was in use [Fischer-Bossert 03]) based a duplex Cu-Ag patination process onto a lead substrate. This technique does dos not belong to any established method for the growth of an artificially silvered coating in classical antiquity. Indeed the gathered evidence clearly indicates that the original appearance of this artefact neither can be explained in terms of: (i) the pyro-metallurgical process utilised to obtain the subærati-type counterfeit coins [La Niece 93], nor can it be reconciled with (ii) the various colouring treatments [Giumlia Mair 01] for silver imitation, of which we are now aware of having been adopted in ancient times.

In the pyrometallurgical process (i), a production method [Bozzini 03] based on diffusion annealing treatments of copper flans wrapped with a thin (250 μm thick) silver foil was in use in order to obtain "stuffed" flans; striking of these "stuffed flans" at ambient temperature ensued. As anticipated in the previous section, the observed recrystallisation textures and the thickness of the applied coating are incompatible with the adoption of this forging technology in the making of the coin. Colouring treatment (ii) for silver imitation that have been fully rationalised in terms of physico-chemical methods, are: (a) the depletion silvering process based on the selective leaching of highly debased Cu-Ag alloys, (a) the tinning process in which a silvery surface was obtained through the formation of a Sn based intermetallic compound and (c) the patination processes based on the inverse segregation phenomenon of eutectoid As-Cu in arsenical bronzes.

Further patination techniques are supposed [Giumlia Mair 01] to have been in use in ancient times to give the impression of solid silver; that are still in need of rediscovery and rationalisation. We can quote, as an example, new evidence [Vlachou 02] which demonstrates a positive correlation between mercury and silver in the outer layer of debased silver coins from the late Roman coinage. These results clearly indicate, for the first time, that an amalgam silvering process could have been utilised for coin counterfeiting in ancient times. Further investigations [Ingo 04] have confirmed these suggestions and moreover have shown that the implementation of this forging method has to be preponed in the very least by about 200 years.

On the basis of the following pieces of evidence:
(i)      impossibility of producing these layers of high melting-point metals onto a low melting-point substrate by a pyrometallurgical approach;
(ii)     SEM morphology exhibiting a bumpy structure of the cauliflower type, which can be obtained only by an electrochemical phenomenon;
(iii)    XRD texture which allows to rule out hammering-and-annealing processes,
we can conclude that the achievement of the observed structure by an electrochemical displacement process is extremely likely

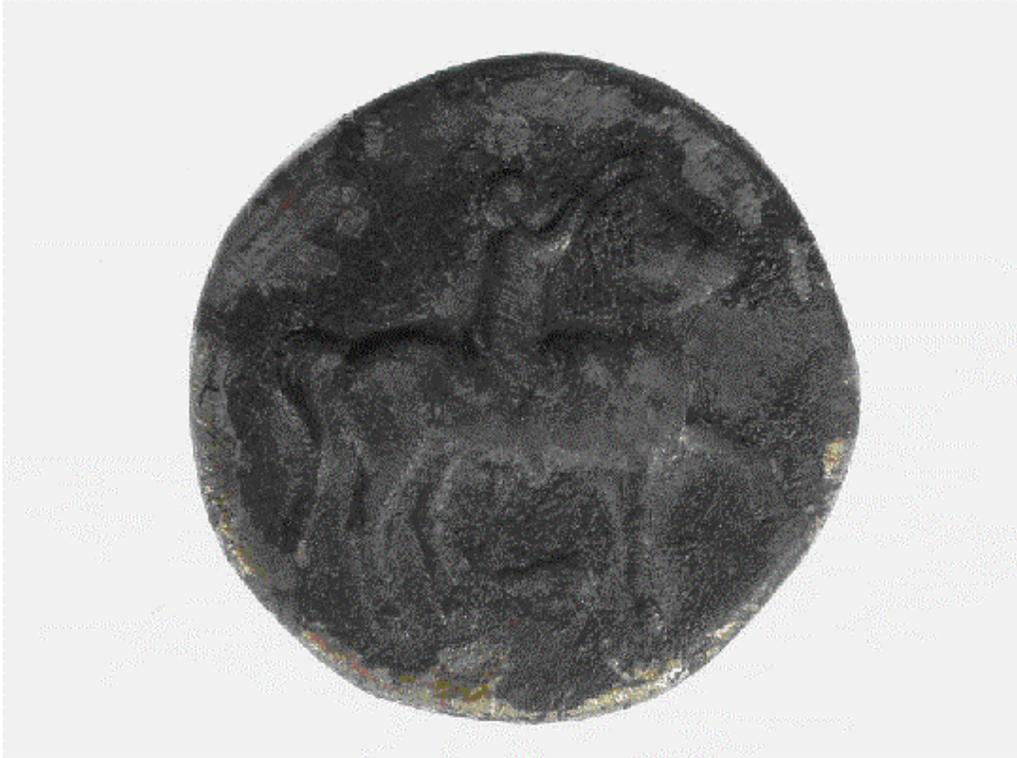

Figure 1 – Obverse: Rider crowning horse walking right; dolphin below.

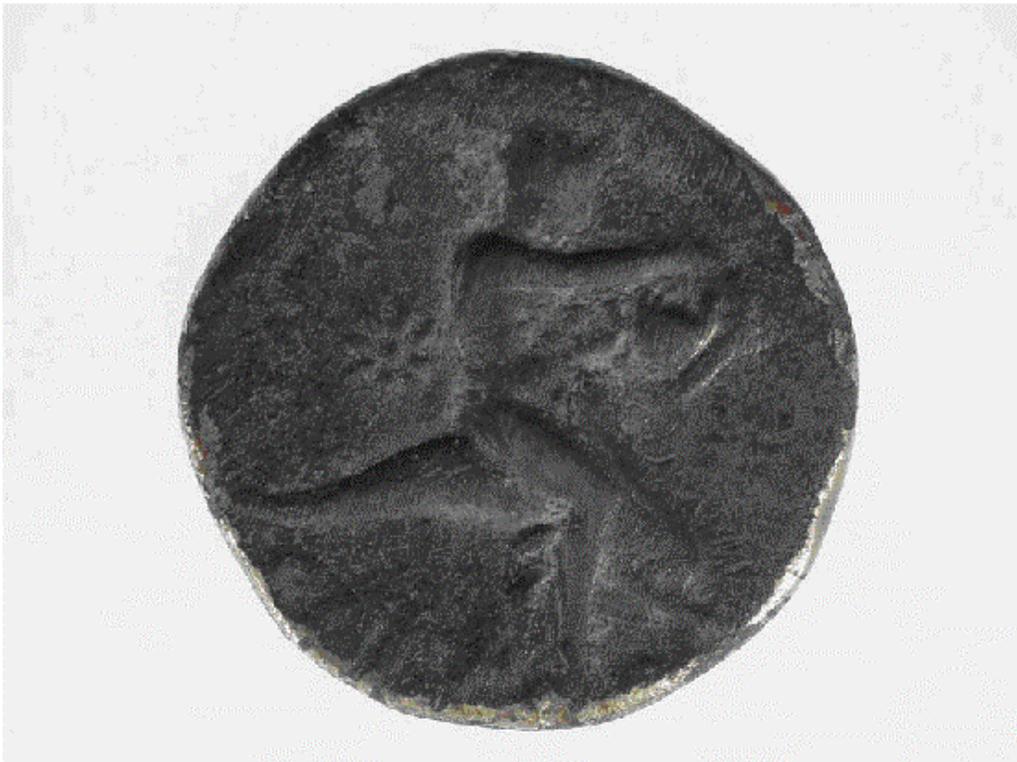

Figure 2 – Reverse: Young man on dolphin right holding phrigian helmet, stars on either side, waves below.

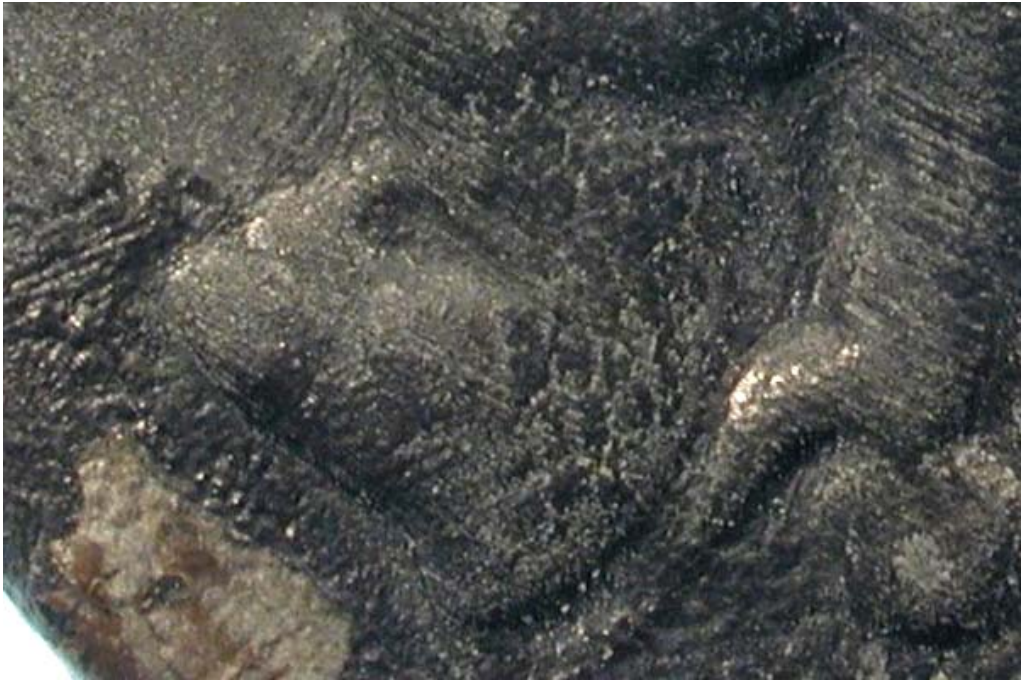

Figure 3 - Uncovered central portion of the coin. Traces of intentional brushing meant to obliterate the solidification flaws.

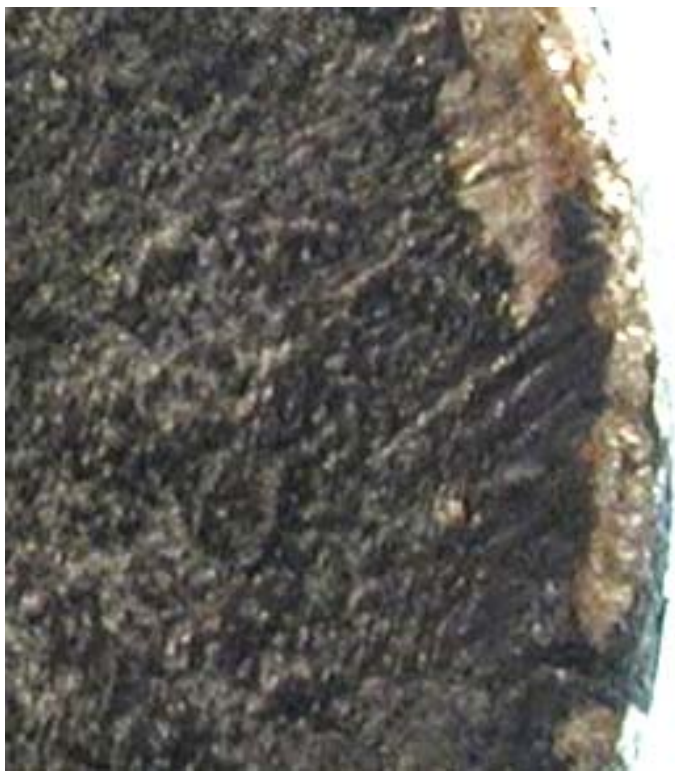

Figure 4 – Traces of intentional brushing, corresponding to solidification flaws. Rim of the coin exhibiting remnants of the coating.

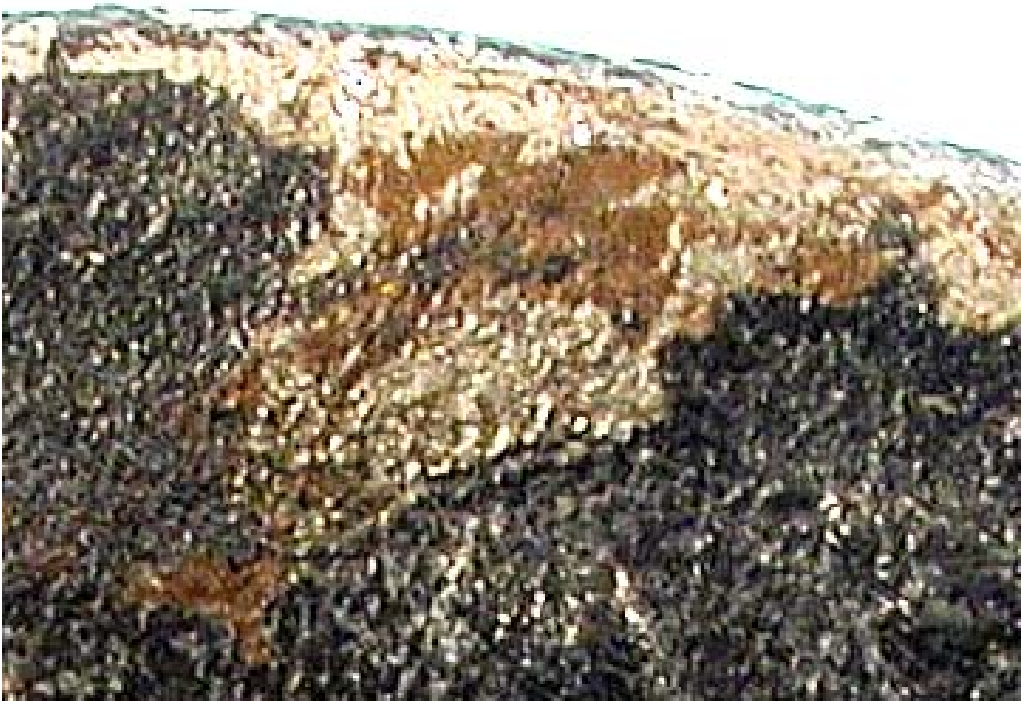

Figure 5 – Macrograph of scratches on the surface of the coin, suggesting that the coating is a bi-layer consisting of different metals.

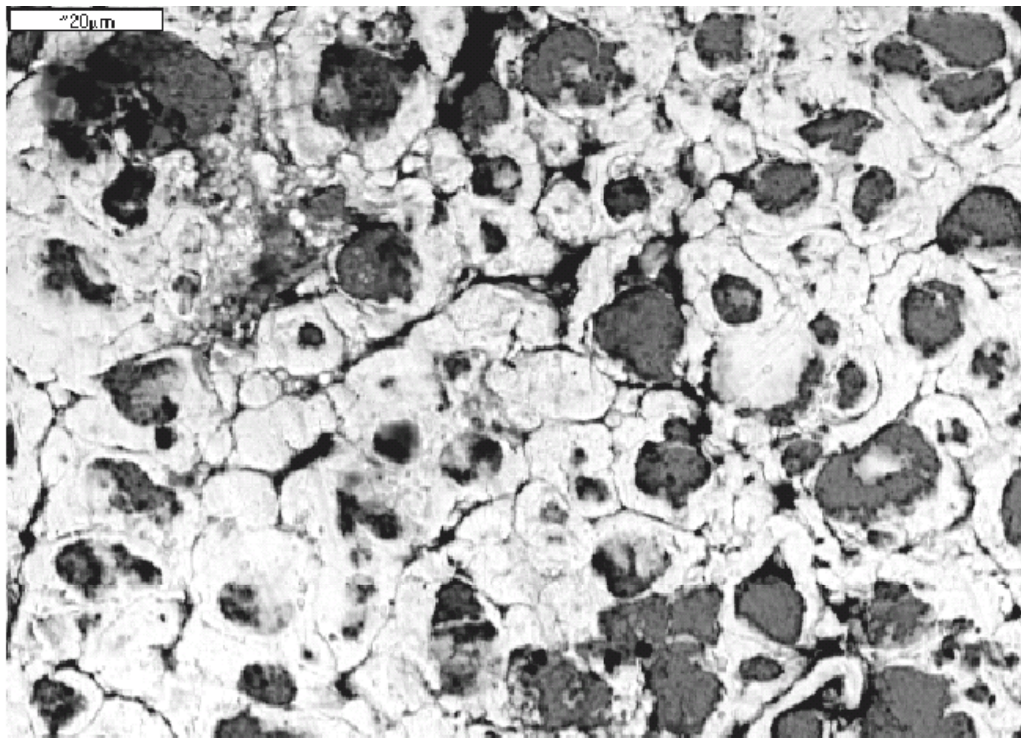

Figure 6 – In-plane SEM micrograph of the residual white-metal coating, exhibiting a "cauliflower" morphology.

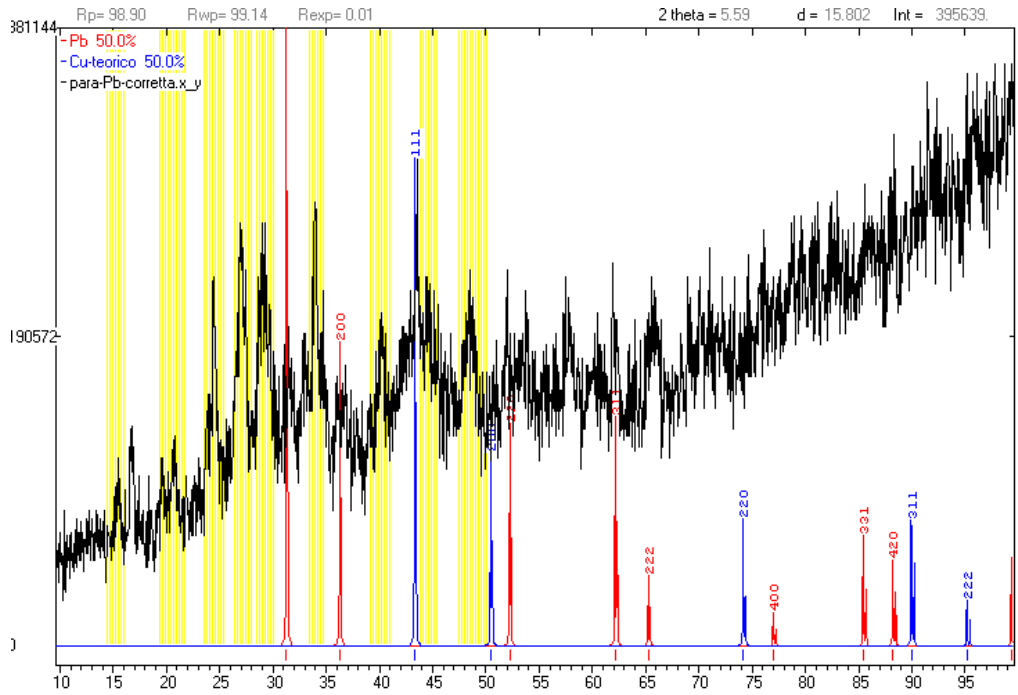

Figure 7 - X-ray diffractogram of the uncovered portion of the coin.

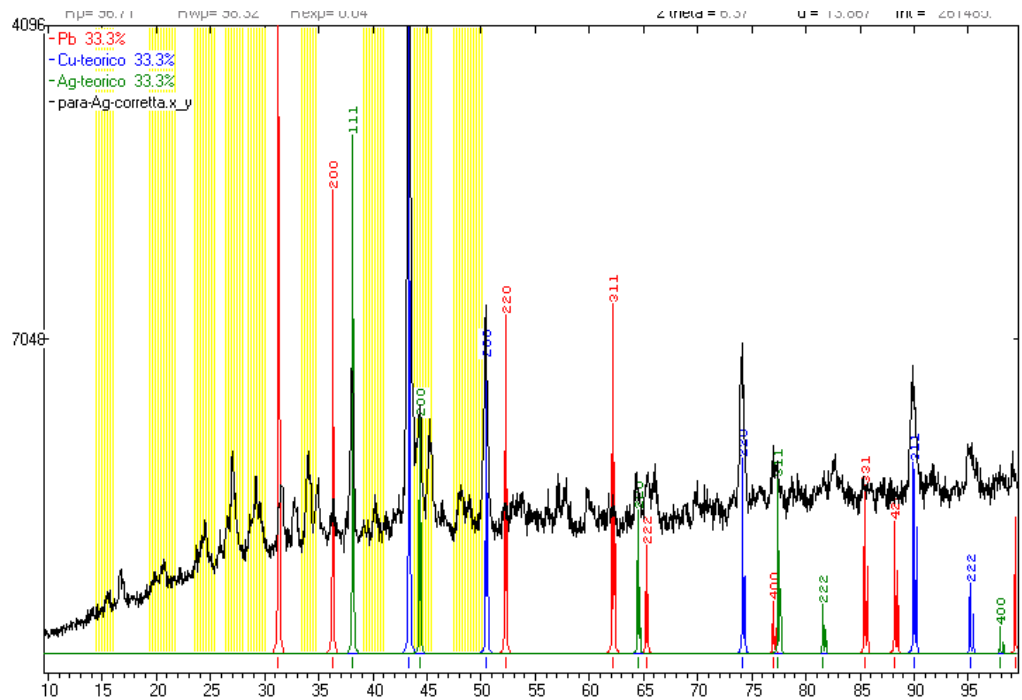

Figure 8 - X-ray diffractogram of the coated portion of the coin.